\documentstyle[12pt, epsfig]{article}

\newcommand{\be}{\begin{equation}}
\newcommand{\ee}{\end{equation}}

 1
\font\elevenrm=cmr10 scaled\magstep 1
 1

\begin{document}
\vspace*{1.8cm}
  \centerline{\bf X RAY PRECURSORS IN GRBs and SGRs: }
  \centerline{\bf OUTER $X$ TAILS AROUND A PRECESSING $\gamma$ JET}
\vspace{1cm}
  \centerline{Daniele Fargion}
\vspace{1.4cm}
  \centerline{Physics Department and INFN, Rome University 1}
  \centerline{\elevenrm Ple.A.Moro 2, 00185, Rome, Italy}
\vspace{3cm}
\begin{abstract}
 Weak isolated X-ray precursor events before the main Gamma Ray Burst, GRB,
and also rare Soft Gamma Repeaters, SGR, events are in
disagreement with any Fireball, or Magnetar, scenarios.
  These models are originated by an unique explosive event leading, by
internal-external shock waves, to softer secondary trains
following a main gamma signals. Indeed the earliest $GRB980519$,
 $GRB981226$ events as well as the latest and most distant
identified one as GRB000131 are showing rare but well identified
and distinct X Ray precursor, occurring tens of seconds or even a
minute before the main GRB eruption. These weak X precursors
bursts correspond to non-negligible energy powers, up to million
Supernova ones.
  They are rare, about $(3-6)\%$ of all GRBs, but not unique.
  Similar huge explosive precursor are in total disagreement with a successive
main Fireball GRB outburst.
  Comparable brief X-ray precursor flashes are found also in rarest and most
detailed SGRs events as those observed on 27 and 29 August 1999
from SGR 1900+14.
  They are inconsistent with a Magnetar Fireball explosion.
  We interpret them as earlier marginal blazing of outlying X conical Jet tails
surrounding a narrower gamma precessing,spinning beamed Jet in
blazing mode toward the Earth; later re-crossing and better
hitting of the target -the satellite detectors- is source of the
main GRB (and SGR) observed structured event.
  The X Ray precursor existence is an additional remarkable evidence of the
Precessing relativistic Jet Nature of both GRBs and SGRs.

\end{abstract}
\vspace{2.0cm}

\section{Introduction: GRBs and SGRs: Spinning, Precessing and  Blazing $\gamma$ Jet}

Gamma Ray Burst and Soft Gamma Repeaters reached an apparent stage
of maturity: tens of GRBs found, finally, an X, optical and (or)
radio transient (the after-glow) identification as well as some
associated host galaxies at cosmic red-shifts (Bloom et all 2000).
New categories of GRBs and SGRs events have been labeled, but
even within these wider updated data no conclusive theory or even
partial understanding seem to solve the old-standing GRB/SGRs
puzzle: the Nature of The GRB-SGR signals. \\

  On the contrary the wider and wider collection of data are
  leading to a schizophrenic attitude in the most popular
  isotropic models, the Largest Cosmic Explosions (Fire-ball, Hypernova,
  Supra-Nova) with more and more phenomenological
  descriptions (power laws everywhere) and less and less unifying
  views. This "give up" attitude seem to reflect the surprising
  never ending morphologies of GRBs.
  \\ We argued on the contrary
  that GRBs and SGRs find a comprehensive theory within a thin spinning
  and multi precessing $\gamma$ Jet, sprayed by a Neutron Star, NS, or Black Hole, BH,
  (Fargion 1994-1999, Fargion, Salis 1995-1998).
  For instance the extreme energy released in last GRBs ($\gg$ $10^{54}$
  $erg $), comparable to few solar masses or more, leads to a deep
   conflict with any isotropic GRB energy (masses and corresponding Schwarchild
   scale times above milliseconds) and the sharp
   observed GRBs fine time structures (below a fraction of millisecond).
     Moreover the energy power spread (from $10^{53}$
  $erg s^{-1}$) for most far GRBs versus $10^{46}$  $erg  s^{-1}$ for
  nearest GRB980425, led most Fireball defenders
   to neglect, hide or even reject in a very arbitrary way
  the  nearest and best identified GRB connection to a Supernova, SN,
  explosion as SN1998bw.
  Finally the same rarity of GRB-SN detections
  and the established GRB980425-SN link favors
   the thin Jet Nature of GRBs.\\

   Even originally ($1970-80$) unified  GRB/SGR models since last fifteen years
   are commonly separated by their repeater and spectra differences;
 however  very recently they openly shared
  the same spectra, time and flux structures.
  This favours once again their common Nature. However they are
  up present times usually described by catastrophic spherical explosions,
   but, as we shall show in this paper, they should not be.
   Their different distances, cosmic versus galactic ones, imply
different power source Jet, but their morphological similarity
strongly suggest an unique process: the blazing of  a spinning and
multi-precessing gamma Jet, from either Neutron Star or Black
Hole. The $\gamma$ Jet is born by high GeVs electron pairs Jet
which are regenerating, via Inverse Compton Scattering, an inner
collimated beamed $\gamma$ (MeVs) precessing  Jet. The thin jet
(an opening  angle inverse of the electron Lorentz factor, a
milli-radiant or below), while spinning, is driven by a companion
and/or an asymmetric accreting disk in a Quasi Periodic
Oscillation (QPO) and in a Keplerian multi-precessing blazing
mode: its $\gamma-X$ ray lighthouse trembling and flashing is the
source of the complex
and wide structure of observed Gamma Bursts. \\
These $\gamma$ Jets share a peak power of a Supernova ($10^{44}
erg s^{-1}$) at their birth (during SN and Neutron Star
formations), decaying by power law $\sim t^{-1}$ $-$ $\sim
t^{-(1.5)}$ to less power-full Jets that converge to present
persistent SGRs stages. Indeed these ones  are blazing events
from late relic X pulsar observable only at nearer distances. The
$\gamma$ Jet emit in general at $\sim$ $ 10^{35}$ erg$ s^{-1}$
powers; both of GRB and SGR show an apparent luminosity amplified
by the inverse square of the thin from $10^{-3}$ to $10^{-4}$
radiant angle Jet beaming: the corresponding solid angle
$\Omega$ spreads between $10^{-7}$ and $10^{-9}$.\\
 Optical-Radio After-Glows are not the fading fireball
explosion tails often observed in puzzling variable non monotonic
decay, but the averaged external Jet tails moving and precessing
and geometrically fading away. The rare optical re-brightening
(the so called SN bump) observed in few afterglow has been
erroneously associated to an underlying isotropic SN flash: it is
more probably the late re-crossing of the precessing Jet periphery
toward the observer direction.

%% ATTENZIONE !!!!
%% In particular the geometrical beaming offered in the rare
%% GRB970508 a peculiar optical re-brightening  and a manifest radio
%% oscillating afterglow
%% by the spinning and precessing Jet. \\

One of the most recent and convincing evidence against any
explosive GRB model, confirming present $\gamma$ precessing Jet
theory, is hidden in the in the recent GRB 000131 data which show
an un-explicable, for Fireball model, X precursor signal 7 sec
long, just  62 seconds prior to the huge main gamma trigger. How
could any GRB source coexist such a powefull precursor?

\begin{figure}[thb]
 \begin{center}
  \mbox{\epsfig{file=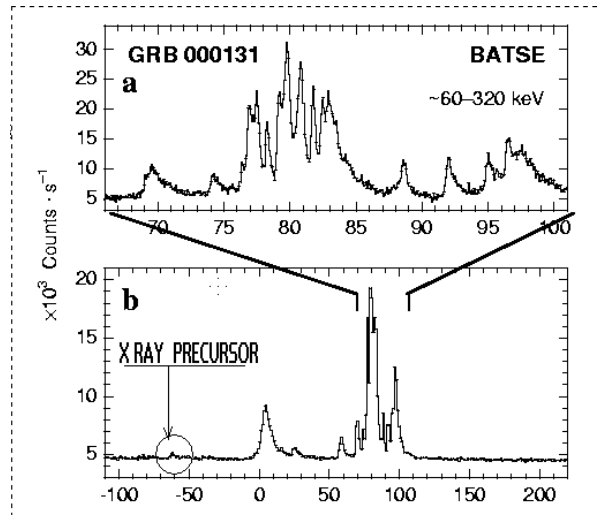, width= 8cm}}
  \caption{\em {Location and Intensity of early $X$ Precursor in GRB000131}}
 \end{center}
\end{figure}

\begin{figure}[thb]
 \begin{center}
  \mbox{\epsfig{file=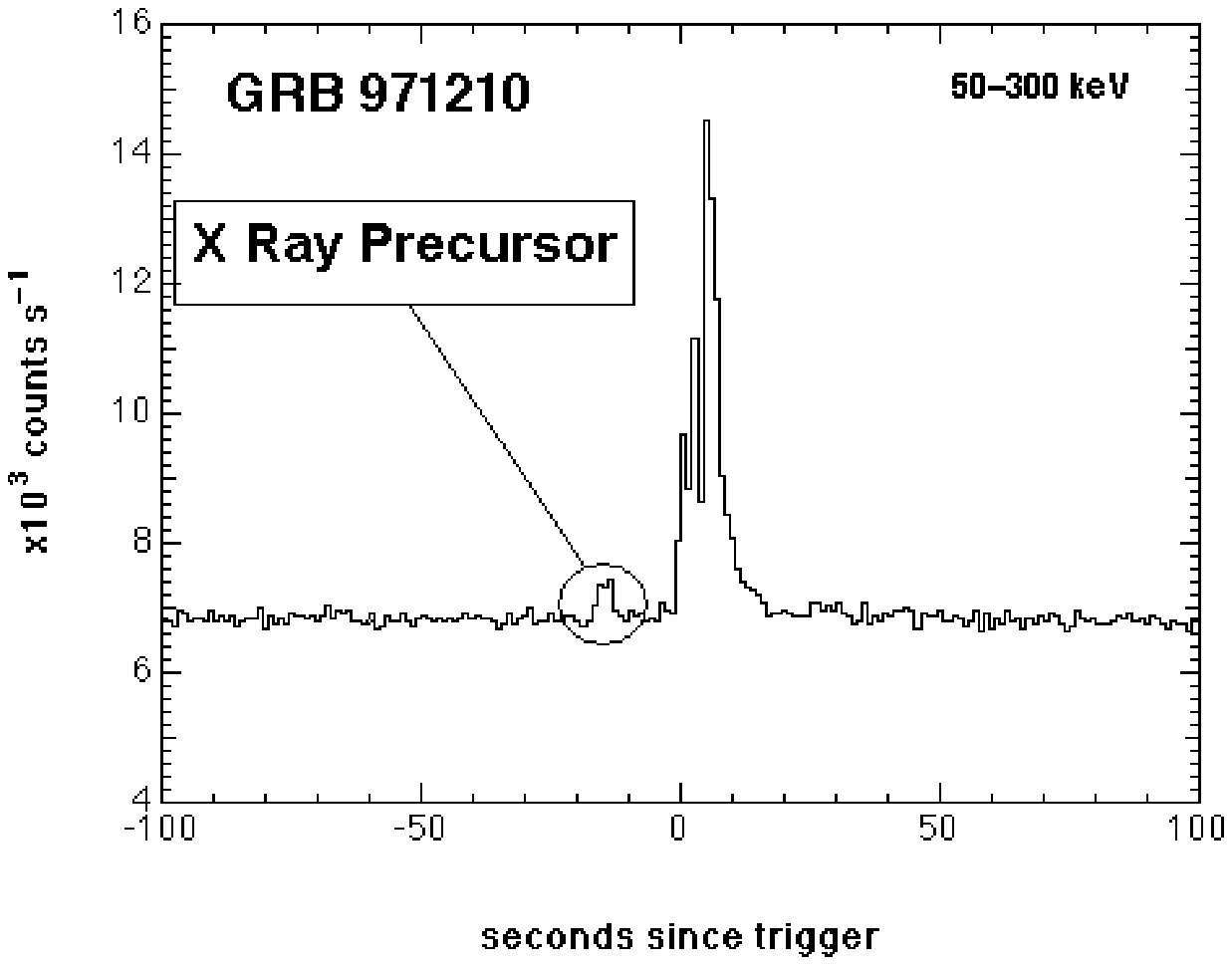, width= 6cm, bb=10 20 400 200}}
  \mbox{\epsfig{file=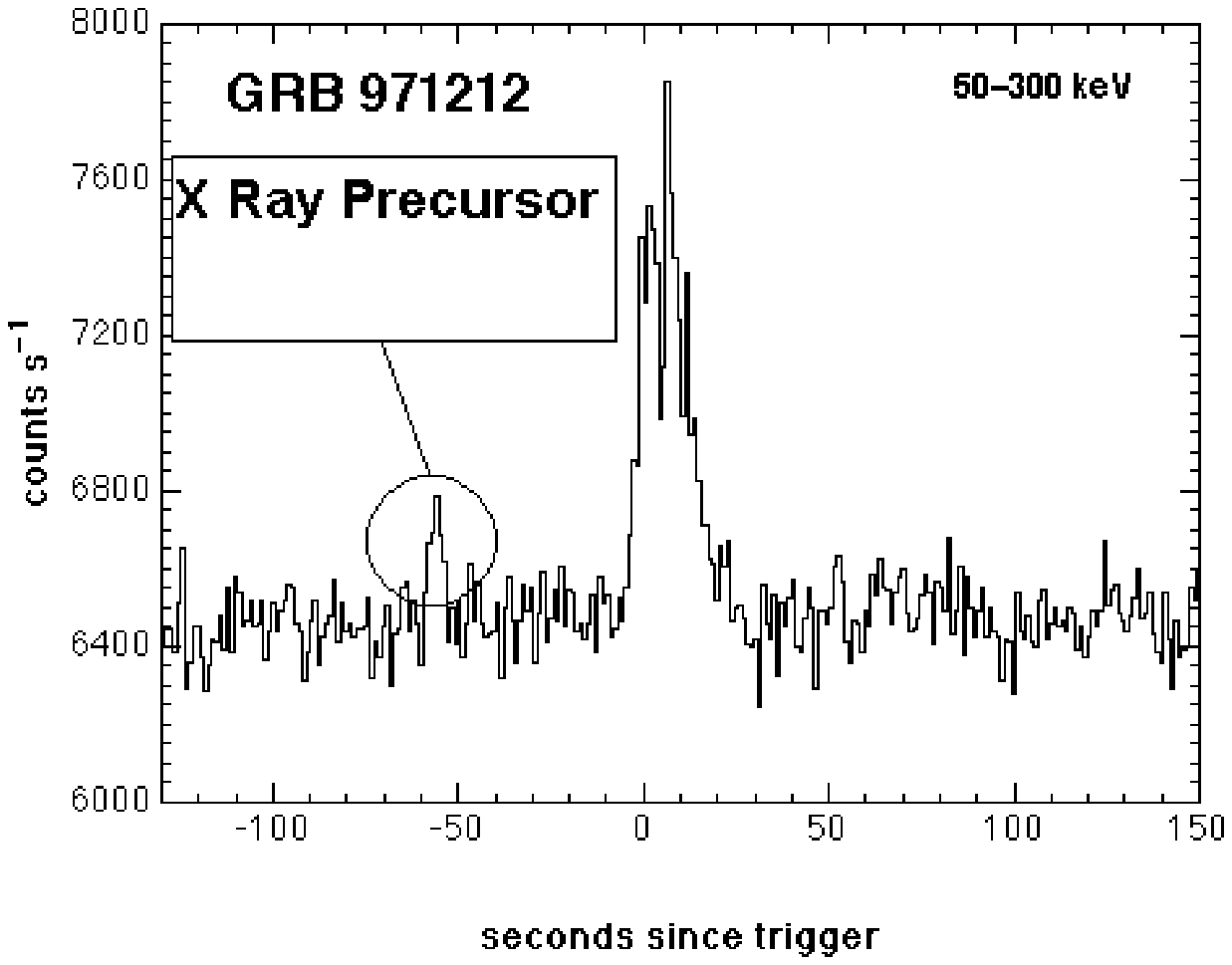, width= 6cm, bb=10 20 400 200}}
  \caption{\em {Fig $2a$ and $2b$: Time evolution and X precursors in GRB $971210$ and GRB $971212$ }}
 \end{center}
\end{figure}

\begin{figure}[thb]
 \begin{center}
  \mbox{\epsfig{file=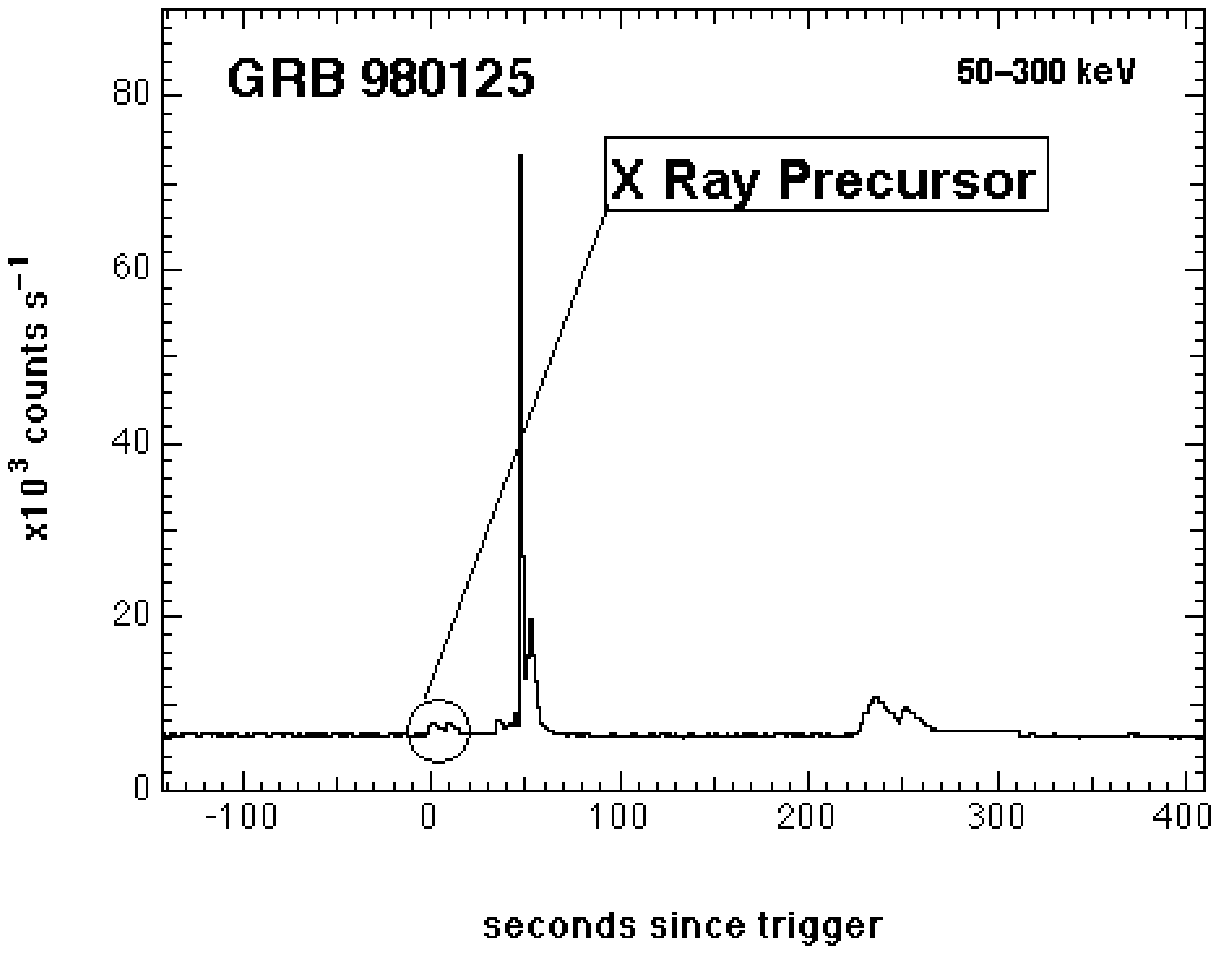, width= 6cm, bb=10 20 400 200}}
  \mbox{\epsfig{file=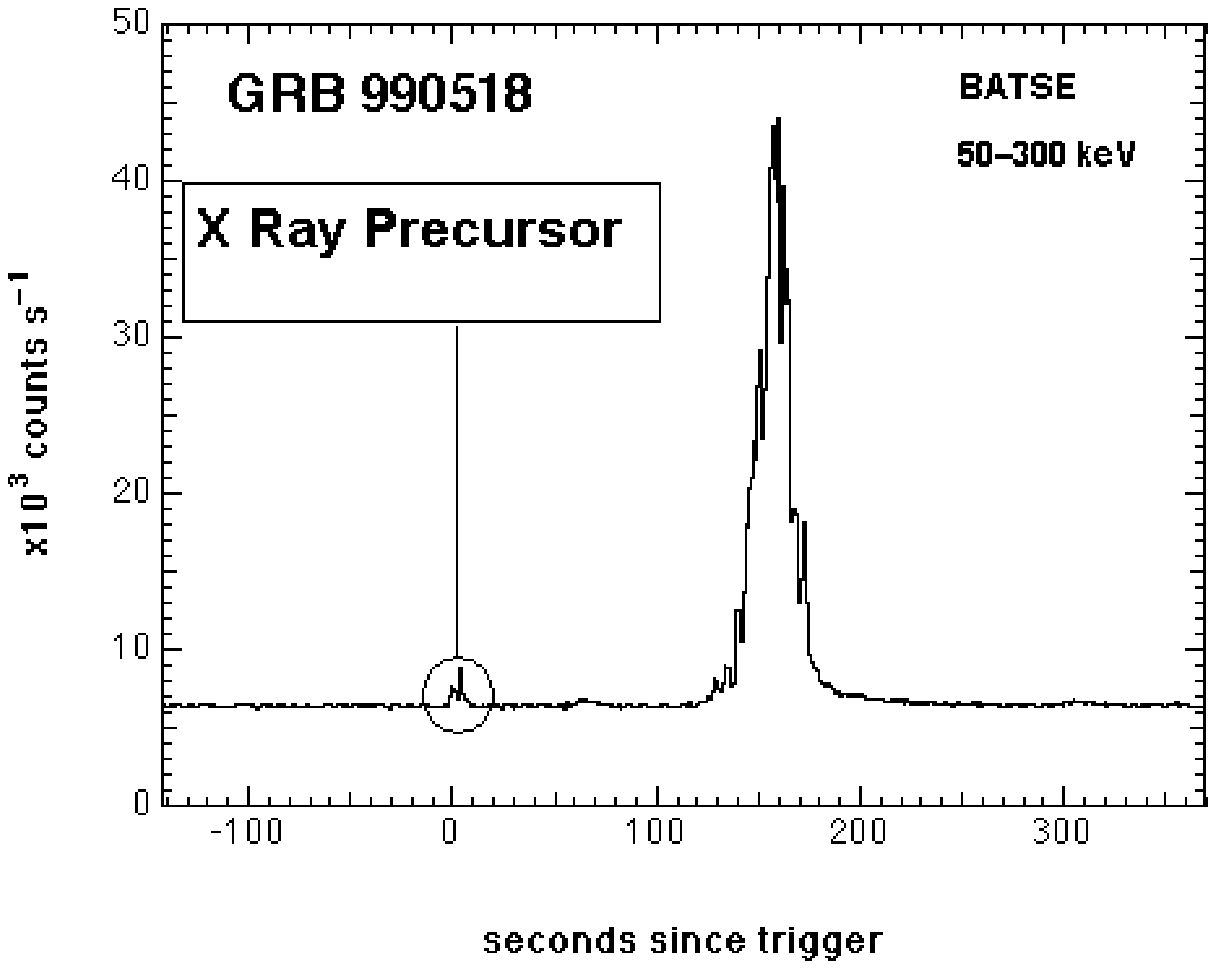, width= 6cm, bb=10 20 400 200}}
  \caption{\em {Fig $3a$ and $3b$: Time evolution and X precursor in GRB $980125$ and GRB $990518$ }}
 \end{center}
\end{figure}

\begin{figure}[thb]
 \begin{center}
  \mbox{\epsfig{file=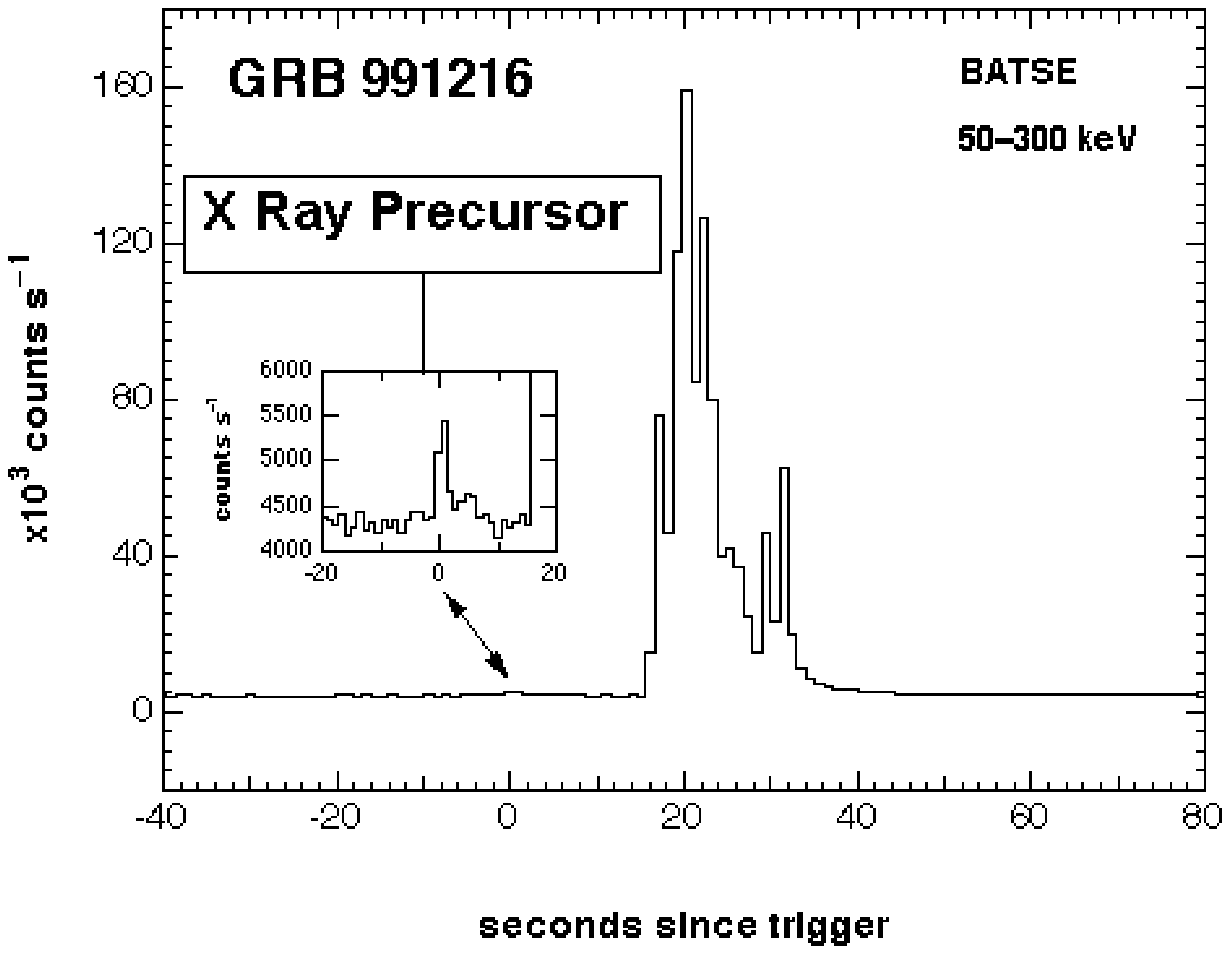, width= 6cm, bb=10 20 400 200}}
  \mbox{\epsfig{file=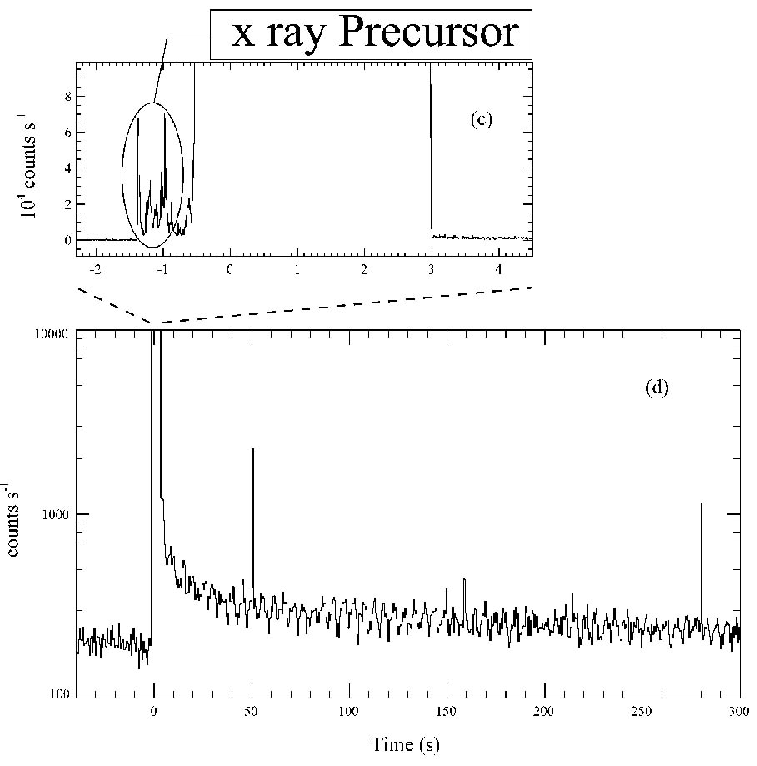, width= 7cm, bb=10 20 300 200}}
  \caption{\em {Fig $4a$ and $4b$: Time evolution and X precursor in GRB $991216$ and in SGR $1900+14$ on 29 August $1998$ }}
 \end{center}
\end{figure}

\section{ Are GRBs an Unique $\gamma$ Jet Explosion?}

  As we noted above  GRBs are not showing any standard candle
  behaviours within any Fireball isotropic model. Some moderate wide
  Fireball-Jet (a not conclusive compromise, like so called collapsar combining SN
  and open fountain-like Jet explosion) models with a  large beaming (ten degree
  opening) can accommodate all cosmic GRBs excluding the "problematic" nearest
  $GRB980425$ event  keeping it out of the frame.\\
  Nevertheless SGRs, which share, some time, the same GRB
  signature, are themselves still within a popular isotropic
  mini-Fireball scenario powered by Magnetar  explosive
  events. This model imply a magnetic energy in the neutron star at least 4
  order of magnitude above the kinetic rotational energy, calling for
  an anomalous and unexplained energy e-qui-partition bias.\\
  Beaming may solve the puzzle within a common X-ray Pulsar power.

    Indeed  the SGR1900+14 event BATSE   trigger 7171 left an almost
     identical event comparable to a just following  GRB (trigger 7172) on
   the same day, same detector, with same spectra and comparable flux.
      This Hard-Soft connection has been re-discovered
      and confirmed   more recently  by BATSE group: ( Fargion 1998-1999;Woods et all
      1999) with an additional hard event of SGR 1900+14 recorded in GRB990110 event.\\
    2)an additional GRB-SGR connection occur between  GRB980706
   event with an almost identical (in time, channel spectra, morphology and
   intensities) observed in GRB980618 originated by SGR 1627-41.
    Nature would be   very perverse in mimic two signals,
     (even  if at different distances and different powers),
    by two extreme different source engines.\\

   To decide for a  model intuitively  let us just consider with no prejudice
   the  last reported (and most distant $z= 4.5$) event:
   GRB000131 and its $X$ ray precursor: \\
   This event while being red-shifted and slowed down by a
   factor 5.5 exhibit on the contrary a short scale time fine structure not explicable
   by any fireball model, but  well compatible  with
    a thin, fast spinning precessing $\gamma$ jet.\\
   The extreme $\gamma$ energy budget, calling for a
   comparable $\nu$ one, exceeds few solar masses in its main
   emission even for ideal full energy conversion.\\
    Moreover one must notice the presence of a weak $X$-ray precursor pulse
   lasting 7 sec, 62 sec before the huge main structured $\gamma$ burst
   trigger. Its arrival direction (within 12 degree error) with main GRB
   is consistent only with the main pulse
    ( a probability to occur by chance below$3.6 10^{-3}$).\\
    The time clustering proximity (one minute over a day GRB rate average) has the probability
   to occur by chance below once over a thousand.
   The over all probability to observe this precursor by change is below 3.4 over a million
   making inseparable its association with the main  GRB000131 event.
    This weak burst signal correspond to a power above a million
   Supernova and have left no trace or Optical/X transient just a minute
   before the real (peak power $> billion $ Supernova) energetic event.
   Similar X precursors occured in a non negligible minor sample of GRBs (see for example Fig
   2-4a) and also few SGRs event (Fig 4b).

   How could any isotropic  GRB explosive progenitor survive such a disruptive isotropic
    precursor trigger? Twice a miracle? Only a persistent precessing Gamma Jet
    crossing nearby the observer direction twice could do it.

\section{The GRB-SN Connection and the Thin Precessing $\gamma$ Jet}

 The need for GRBs beaming is wide: the GRB luminosities are
 over Eddington, the event peaked structure is chaotic, the spectra
 is non-thermal, the energy budget may exceed two solar masses
 annihilation (Fargion 1994, Fargion, Salis 1995-98, Fargion 1998-1999).
  The spinning and precessing periodicity is hidden into the
 short GRB observational window;  indeed the periodicity did finally arise
  in Soft Gamma Repeaters as soon as  more data have been  available.
  As it was demonstrated recently,
 many light curves  and spectra of the GRB might be explained by the blazing of
multi-precessing   gamma jets (Fargion 1994-1995-1999).
 A wider GRBs data sheet, as  for SGRs data
  would show the spinning periodicity of GRBs and possibly the quasi
  periodic oscillations (QPO) of the parental  binary system.\\

 Behind the energy problem stand (to isotropic fireball models)
 the puzzling low probability to observe
any  close GRB as GRB980425 at a negligible cosmic distance (38
Mpc) along with a couple of dozen of far and very far events seen
by  BeppoSax in last two years.

Statistical arguments (Fargion 1998, 1999) favor a unified GRBs
model based on blazing, spinning and precessing thin jet. We
assume that GRB jet arise in most SNe outbursts. The far GRBs are
observables at their peak intensities (coincident to SN) while
blazing in axis to us within the thin jet very rarely;
consequently the hit of the target occurs only within a wide
sample of sources found in  a huge cosmic volume. In this frame
work the GRB rate do not differ much from the SN rate. Assuming a
SN-GRB event every 30 years in a galaxy and assuming a thin
 angular cone ($\Omega < $(1/4)$10^{-8}$) the probability to be within
 the cone jet in a ($ 4*10^{10}$)cosmic  sample of galaxies
  (at limited Hubble R$\geq$ 28 magnitude ) within
 our main present observable Universe volume ($z \sim 1$-$z \sim 4$)
 during one day of observation at a nominal 10 sec GRB
 duration is quite small: ($P < 10^{-2}$). This value should be suppressed
 by nearly an order of magnitude because of the detector acceptance. However a precessing
 gamma jet whose decaying scale time is a thousand time longer than the GRB itself
 (decaying by a power law $\sim t^{-1}$)
 whose scale time is  nearly  ten or twenty thousand of seconds,
  may fit naturally the observed  GRB rate.\\

Also, if these jets have complicated spinning and
multi-precession spirals, they could explain many (or all)
features  of the light-curves of GRB. Consider especially the
observed periodic tails in SGR signals  and rarest (3\%-6\%)
mini-X-GRB precursors: their periodicity has been discovered only
when long data sample of  events have been collected, contrary to
GRBs where the decaying power and the distance make difficult
recording their
signal tails.\\
\section{The Spiral Jets and the Rings in SN1987A}
 The precessing Jet signature is hidden in different forms.
 The possibility that precessing Gamma jets are source by
 their  interactions onto a red giant relic shell
 of the Twin Ring around SN1987A  has been proposed since 1994
 (Fargion \& Salis 1995b, 1995c). It has been also
been suggested (Fargion \& Salis 1995) that the variable
presence  of a paraboloid thin arc  along one of the twin ring of
SN1987A, the mysterious "Napoleon Hat" observed on 1989-1990, was
the  evidence for a thin long projected jet interacting tens
parsec away from the SN1987A toward us. The jet pressure could
also accumulate gas and form dense filamentary gas. Such gas
filament fragment as well as  gravitationally clusterize  may lead
to contemporaneous  stellar arc formations.(Y.Efremov\&D.Fargion,
2000).

\section{GRBs and their Neutrino $\nu$ and $\gamma$\\
 Energy Budget in Fireball}

It is  surprising (at least to the author) that after a decade of
fireball inflation papers, at present crisis (GRB990123,
GRB990510, GRB991226, GRB000131 over energetic event) there is no
definitive rejection of this popular isotropic model. On the
contrary there is wide spread resistance and inertia to give up
with some excuse to this famous  misleading fireball model.\\

Gamma Ray Bursts as recent $GRB990123 and GRB990510$ emit, for
isotropic explosions, $\gamma$ energies as large as  two solar
masses annihilation. These energies are underestimated because of
the neglected role of comparable ejected MeV (Comptel signal)
neutrinos $\nu$ bursts and assume an unrealistic ideal energy
conversion efficiency. Indeed, as often neglected, it is
important to remind that the huge energy bath (for a fireball
model) on GRB990123 imply also a corresponding neutrino burst. As
in hot universe, if entropy conservation holds, the $\nu$ energy
density factor to be added to the photon $\gamma$  budget is at
least $( \simeq (21/8)\times (4 /11)^{4/3} )$. If entropy
conservation do not hold the energy needed is at least a factor
$[21/8]$ larger than the gamma one. The consequent total
energy-mass needed for the two cases are respectively 3.5 and 7.2
solar masses. Additional factors must be introduced for realistic
energy conversion efficiency leading to energies as large as tens
of solar masses. No fireball
by NS may coexist with it. Jet could.\\
 These extreme power cannot be explained with any
standard spherically symmetric Black Hole Fireball model. A too
heavy Black Hole (hundred or thousands solar masses) or, worse,
Star would be unable to coexist with the shortest millisecond time
structure of Gamma Ray Burst. Cosmological and nearby
Gravitational Red-shifts may only make the Fireball Model more
inconsistent. Smaller size BH or NS do not offer enough mass
reserve. Beaming of the gamma radiation may overcome the energy
puzzle along with the short scale-time. However any mild
''explosive beam'' event as some models (Wang \& Wheeler 1998)
$(\Omega
> 10^{-2} )$ would not solve the jet containment at the
corresponding disruptive energies. Moreover such a small beaming
would not solve the huge GRBs flux energy windows ($10^{47} \div
10^{54}$ erg/sec), keeping GRB980425 and GRB990123 within the
same GRB framework.

 Only extreme beaming $(\Omega \sim 10^{-8} )$, by a
slow decaying, but long-lived precessing jet, may coexist with
characteristic Supernova energies, apparent GRBs output and the
puzzling GRB980425 statistics as well as the GRB connection with
older, nearer and weaker SGRs relics. Therefore SGRs are very
useful nearby astrophysical Laboratory where to study and test
the far GRB process. SGRs are not associated with huge OT
afterglow or explosive SN event. Indeed they are persistent Jet
eventually feed by an accreting companion or disk.
 The optical transient OT of GRB is in part
due to the coeval SN-like explosive birth of the jet related to
its maximal intensity; the OT is absent in older relic Gamma jets,
the SGRs because their time-scales are too short (few seconds or
below)  to being revealed. Their explosive memory is left around
in their relic nebula or plerion injected by the Gamma Jet which
is running away. The late GRB OT, days after the burst, are
related mainly to the Jet tail precession; it is usually enhanced
only by a partial beaming $(\Omega \simeq 10^{-2} )$. The extreme
peak OT during GRB990123 (at a million time a Supernova
luminosity) is just the extreme beamed $(\Omega \leq 10^{-5} )$
Inverse Compton optical tail, responsible of the same extreme
gamma (MeV) extreme beamed
$(\Omega \leq 10^{-8})$ signal.\\

\section{The Peculiar Nearest, Weakest, Slowest\\
GRB980425 in SN1998bw}

 Finally Fireballs are unable to explain the following
  key questions (Fargion 1998-1999)
related to the association GRB980425 and SN1998bw (Galama et
all1998):

\begin{enumerate}
 \item Why nearest ``local'' GRB980425 in ESO 184-G82 galaxy at
 redshift $z_2 = 0.0083$ (nearly 38 Mpc.) and the most far away ``cosmic'' ones as
 GRB971214 (Kulkarni et al.1998) ( or GRB00131 (Andersen et al.
 2000))  at redshift $z_2 = 3.42$ (and  $z_2= 4.5$)
 exhibit a huge average and peak
 intrinsic luminosity ratio?
\begin{equation}\label{eq1}
\frac{<L_{1 \gamma}>}{<L_{2 \gamma}>}  \cong  \frac{<l_{1
\gamma}>}{<l_{2 \gamma}>} \frac{z_{1 }^2}{z_{2 }^2} \cong 2 \cdot
10^5 \;\;; \left. \frac{L_{1 \gamma}}{L_{2 \gamma}} \right|_{peak}
\simeq 10^7.
\end{equation}
Fluence ratios $E_1 / E_2$ are also extreme ($\geq 4 \cdot 10^5$).
 \item Why GRB980425 nearest event spectrum is softer than cosmic
 GRB971214 while Hubble expansion would naturally imply the opposite by a
 redshift factor $(1+z_1)\sim 4.43$?
 \item Why, GRB980425 time structure is
 slower and smoother than cosmic one, as above contrary to Hubble law?
 \item Why we observed so many (even just the rare April one over 14
Beppo Sax optical transient event) nearby GRBs? Their probability
to occur, with respect to a cosmic redshift  $z_1 \sim 3.42$ must
be suppressed by a severe volume factor
\begin{equation}\label{eq2}
\frac{P_{1}}{P_{2}} \cong \frac{z_{1}^{3}}{z_{2}^{3}} \simeq 7
\cdot 10^{7} \;\;\;.
\end{equation}
\end{enumerate}
The above questions remain unanswered by fireball candle model.
Indeed hard defenders of fireball models either ignore the problem
or, worse, they negate the same reality of the April GRB event. A
family of new GRB fireballs are ad hoc and fine-tuned solutions.
We believed since 1993 (Fargion 1994) that spectral and time
evolution of GRB are made up blazing beam gamma jet GJ. The GJ is
born by ICS of ultrarelativistic (1 GeV-tens GeV) electrons
(pairs) on source IR, or diffused companion IR, BBR photons
(Fargion, Salis 1998). The beamed electron jet pairs will produce
a coaxial gamma jet. The simplest solution to solve the GRBs
energetic crisis (as GRB990123 whose isotropic budget requires an
energy above two solar masses) finds solution in a geometrical
enhancement by the jet thin beam.\\

\section{Hard Gamma Jet by Inverse Compton Scattering of GeV Electron Pairs}

 A jet angle related by a
relativistic kinematics would imply $\theta \sim
\frac{1}{\gamma_e}$, where $\gamma_e$ is found to reach $\gamma_e
\simeq 10^3 \div 10^4$ (Fargion 1994, 1998). At first
approximation the gamma constrains is given by Inverse Compton
relation: $< \epsilon_\gamma > \simeq \gamma_e^2 \, k T$ for $kT
\simeq 10^{-3}-10^{-1}\, eV$ and $E_e \sim GeVs$ leading to
characteristic
X-$\gamma$ GRB spectra. \\
The origin of $GeVs$ electron pairs are  very probably decayed
secondary related to primary inner muon pairs jets, able to cross
dense stellar target.\\ However an impulsive unique GRB jet burst
(Wang \& Wheeler 1998) increases the apparent luminosity by
$\frac{4 \pi}{\theta^2} \sim 10^7 \div 10^9$ but face a severe
probability puzzle due to the rarity to observe even a most
frequent SN burst jet pointing in line toward us. Vice-versa one
must assume a high rate of GRB events ($\geq 10^5$ a day larger
even than expected SN one a day). As we noted most authors today
are in a compromise: they believe acceptable only mild beaming
($\Omega> \sim 10^{-3}$), taking GRB980425 out of the GRB
''basket''.

On the contrary we considered
 GRBs and SGRs as multi-precessing
and spinning Gamma Jets and the GRB980425 an off-axis classical
jet. In particular we considered (Fargion 1998) an unique scenario
where primordial GRB jets decaying in hundred and thousand years
become the observable nearby SGRs. Sometimes accretion binary
systems may increase the SGRs activity. The ICS for monochromatic
electrons on BBR leads to a coaxial gamma jet spectrum(Fargion \&
Salis 1995, 1996, 1998): $\frac{dN_{1}}{dt_{1}\, d\epsilon_{1}\,
d\Omega _{1}}$ is
\begin{equation}
\epsilon _{1}\ln \left[ \frac{1-\exp \left( \frac{-\epsilon
_{1}(1-\beta \cos \theta _{1})}{k_{B}\, T\, (1-\beta )}\right)
}{1-\exp \left( \frac{-\epsilon _{1}(1-\beta \cos \theta
_{1})}{k_{B}\, T\, (1+\beta )}\right) }\right] \left[ 1+\left(
\frac{\cos \theta _{1}-\beta }{1-\beta \cos \theta _{1}}\right)
^{2}\right] \label{eq3}
\end{equation}
scaled by a proportional factor $A_1$ related to the electron jet
intensity. The adimensional photon number rate (Fargion \& Salis
1996) as a function of the observational angle $\theta_1$
responsible for peak luminosity (eq. \ref{eq1}) becomes
\begin{equation}
\frac{\left( \frac{dN_{1}}{dt_{1}\, d\theta _{1}}\right) _{\theta
_{1}(t)}}{ \left( \frac{dN_{1}}{dt_{1}\, d\theta _{1}}\right)
_{\theta _{1}=0}}\simeq \frac{1+\gamma ^{4}\, \theta
_{1}^{4}(t)}{[1+\gamma ^{2}\, \theta _{1}^{2}(t)]^{4}}\, \theta
_{1}\approx \frac{1}{(\theta _{1})^{3}} \;\;.\label{eq4}
\end{equation}
The total fluence at minimal impact angle $\theta_{1 m}$
responsible for the average luminosity (eq. \ref{eq1}) is
\begin{equation}
\frac{dN_{1}}{dt_{1}}(\theta _{1m})\simeq \int_{\theta
_{1m}}^{\infty }\frac{ 1+\gamma ^{4}\, \theta _{1}^{4}}{[1+\gamma
^{2}\, \theta _{1}^{2}]^{4}} \, \theta _{1}\, d\theta _{1}\simeq
\frac{1}{(\, \theta _{1m})^{2}}\;\;\;. \label{eq5}
\end{equation}
These spectra fit GRBs observed ones (Fargion \& Salis 1995).
Assuming a beam jet intensity $I_1$ comparable with maximal SN
luminosity, $I_1 \simeq 10^{45}\;erg\, s^{-1}$, and replacing
this value in adimensional $A_1$ in equation \ref{eq3} we find a
maximal apparent GRB power for beaming angles $10^{-3} \div
3\times 10^{-5}$, $P \simeq 4 \pi I_1 \theta^{-2} \simeq 10^{52}
\div 10^{55} erg \, s^{-1}$ within observed ones. We also assume a
power law jet time decay as follows
\begin{equation}\label{eq6}
  I_{jet} = I_1 \left(\frac{t}{t_0} \right)^{-\alpha} \simeq
  10^{45} \left(\frac{t}{3 \cdot 10^4 s} \right)^{-1} \; erg \,
  s^{-1}
\end{equation}
where ($\alpha \simeq 1$) able to reach, at 1000 years time
scales, the present known galactic microjet (as SS433)
intensities powers: $I_{jet} \simeq 10^{39}\;erg\, s^{-1}$. We
used the model to evaluate if April precessing jet might hit us
once again. It should be noted that a steady angular velocity
would imply an intensity variability ($I \sim \theta^{-2} \sim
t^{-2}$) corresponding to some of the earliest afterglow decay
law.

\section{The GRB980425-SN 1998bw Conenction\\
and the Probable GRB980712 Repeater}

Therefore the key answers to the above puzzles (1-4)  are: the
GRB980425 has been observed off-axis by a cone angle wider than
$\frac{1}{\gamma}$ thin jet  by a factor $a_2 \sim 500$
(Fargion1998), $\theta \sim \frac{500}{10^4} \approx \frac{5
\cdot 57^0}{100} \approx 2.85^0 \left( \frac{\gamma}{10^4}
\right)^{-1}$, and therefore one observed only the ``softer''
cone jet tail whose spectrum is softer and whose time structure is
slower (larger impact parameter angle). A simple statistics
favored a repeater hit. Indeed GRB980430 trigger 6715 was within
$4 \sigma$ and particularly in GRB980712 trigger 6917 was within
$1.6 \sigma$ angle away from the April event direction. An
additional event 15 hours later, trigger 6918, repeated making
the combined probability to occur quite rare ($\leq 10^{-3}$).
Because the July event has been sharper in times ($\sim 4 \, s $)
than the April one ($\sim 20 \, s $), the July impact angle had a
smaller factor $a_3 \simeq 100$. This value is well compatible
with the
 expected peak-average luminosity flux evolution in eq.(6, 4): \\

$\frac{L_{04\, \gamma}}{L_{07\, \gamma}} \simeq \frac{I_2\,
\theta_2^{-3}}{ I_3\, \theta_3^{-3}} \simeq \left( \frac{t_3}{t_2}
\right)^{-\alpha} \, \left( \frac{a_2}{a_3} \right)^{\, 3} \leq
3.5$ where $t_3 \sim 78 \; day$ while $t_2 \sim
2 \cdot 10^5 \, s$. \\
The predicted fluence is also comparable with the observed ones
$\frac{N_{04}}{N_{07}} \simeq \frac{<L_{04\, \gamma}>}{<L_{07\,
\gamma}>} \, \frac{\Delta \tau_{04}}{\Delta \tau_{07}} \simeq
\left( \frac{t_3}{t_2} \right)^{-\alpha} \, \left( \frac{a_2}{a_3}
\right)^2
\, \frac{\Delta \tau_{04} }{\Delta \tau_{07}} \geq 3$.\\

\section{The SGRs Hard Spectra and their GRB Link by Precessing Jet}

Last SGR1900+14 (May-August 1998) events and SGR1627-41
(June-October 1998) events did exhibit at peak intensities hard
spectra comparable with classical GRBs. We imagine their nature as
the late stages of jets fueled by a disk or a companion (WD, NS)
star. Their binary angular velocity $\omega_b$ reflects the beam
evolution $\theta_1(t) = \sqrt{\theta_{1 m}^2 + (\omega_b t)^2}$
or more generally a multi-precessing angle $\theta_1(t)$ (Fargion
\& Salis 1996):

\begin{equation}\label{eq7}
  \theta_1(t) = \sqrt{\theta_{x}^2 +\theta_{y}^2 }
\end{equation}

\begin{equation}\label{eq8}
  \theta_{x}(t) =                               %\theta_{1 m}+
  \theta_{b} sin(\omega_{b} t + \varphi_{b})+
  \theta_{psr}sin(\omega_{psr} t)+
  \theta_{N}sin(\omega_{N} t  + \varphi_{N})
\end{equation}

\begin{equation}\label{eq8}
  \theta_{y}(t) = \theta_{1 m}+
  \theta_{b} cos(\omega_{b} t + \varphi_{b})+
  \theta_{psr} cos(\omega_{psr} t)+
  \theta_{N} cos(\omega_{N} t  + \varphi_{N})
\end{equation}
where $\theta_{1 m}$ is the minimal angle impact parameter of the
jet toward the observer, $\theta_{b}$, $\theta_{psr}$,
$\theta_{N}$ are, in the order, the maximal opening precessing
angles due to the binary, spinning pulsar, nutation mode of the
jet axis.

%The $\omega_{b}$, $\omega_{psr}$, $\omega_{N}$ are the
The angular velocities combined in the multi-precession keep
memory of the pulsar jet spin ($\omega_{psr}$), the precession by
the binary $\omega_b$ and an additional nutation due to inertial
momentum anisotropies or beam-accretion disk torques
($\omega_N$). On average, from eq.(5) the $\gamma$ flux and the
$X$ optical afterglow decays, in first approximation, as $t^{-2}$;
the complicated spinning and precessing jet blazing is responsible
for the wide morphology of GRBs and SGRs as well as their
internal periodicity.(See figure 5). Similar descriptions with
more parameters and with a rapid time evolution
of the jet has been also proposed by (Portegies Zwart et all 1999).\\

We predicted  (Fargion et all 1995-1999) that such  relic jet
source to be found in the South-East region of SN1987A where
should be hidden a fast running pulsar accelerated by an off-axis
bent beaming. Jets may propel and inflate plerions as the
observed ones near SRG1647-21 and SRG1806-20. Optical nebula
NGC6543 (``Cat Eye'') and its thin jets fingers (as Eta Carina
ones), the  double cones sections in Egg Nebula CRL2688 are the
most detailed and spectacular lateral view of such jets ''alive''.
Their blazing in-axis would appear in our galaxy as SGRs or, at
maximal power at their SN birth at cosmic edges, as GRBs.

\section{The Morphology of Precessing Jet  Relics}
  The Gamma Jet progenitor of the GRB
   is leaving a trace in the space: usually a nebulae
where  the nearby  ISM  may record the jet  sweeping as a
     three dimensional screen. The outcomes  maybe either
      a twin ring as recent SN1987A has shown, or helix traces as
       the Cat Eye Nebula or more structured shapes as plerions
        and hourglass nebulae.
        How can we explain within an unique jet model such a wide
         diversity?\\

         We imagine the jet as born by a binary system (or
         by an asymmetric disk accreting interaction)
         where the compact companion (Bh or NS) is the source of the
         ultra relativistic electron pair jet (at tens GeV.
       Inverse Compton Scattering on IR thermal photons will produce a collinear
        gamma jet at MeV). The rarest case where the jet
         is spinning and nearly isolated would produce a jet train
         whose trace are star chains as the Herbig Haro ones
         (Fargion, Salis 1995). When the jet
         is modified  by the magnetic field torque of the
         binary companion field the result may be a more rich cone shape.
         If the ecliptic  lay on the same plane orthogonal to the jet
         in an ideal circular orbit than the bending
         will produce an ideal twin precessing cones which is
           reflected in an ideal twin rings (Fargion, Salis 1995).
         If the companion is in eccentric orbit the resultant
         conical jet will be more deflected at perihelion while
          remain nearly undeflected at a aphelion.
          The consequent off-axis cones will play the role of a
          mild "rowing" acceleration  able to move the system
          and speed it far from its original birth (explosive)
          place. Possible traces are the asymmetric external twin
          rings painted onto the spherical relic shell by SN1987a.
          Fast relics NS may be speeded by this processes (Fargion, Salis
          1995a, 1995b, 1995c). Because of momentum conservation this asymmetric
            rowing is the source of a motion of the jet relic  in the
            South-East direction. In extreme eccentric system
            the internal region of the ring are more  powered
            by the nearby encounter leading to the apparent gas arcs.
            If the system is orbiting in a plane different from
              the one orthogonal to the jet the outcoming
              precessing jet may spread into a mobile twin cone
              whose filling may appear as a full cone or a twin hourglass
              by
               a common plerion shape. At late times there is also  possible
               apparent spherical shapes sprayed and structured by a chaotic
               helix.
               External  ISM distribution may also play a role enhancing
                 some sides or regions of the arcs.
              The integral jet in  long times may mimic even
                spherical envelopes but internal detailed inspection
                might reveal the thin jet origin (as in recent Eta
                Carina string jets). Variable nebulae behaviours recently
                observed are confirming our present scenario.

\begin{figure}[thb]
 \begin{center}
%  \mbox{\epsfig{file=GRBs-SGR.eps, width= 12cm}}
%  \caption{\em {Left: Two different Spinning, Precessing Gamma Jet blazing to detector at origin.
%                Right: The consequent X, Gamma time evolution signals}}
  \mbox{\epsfig{file=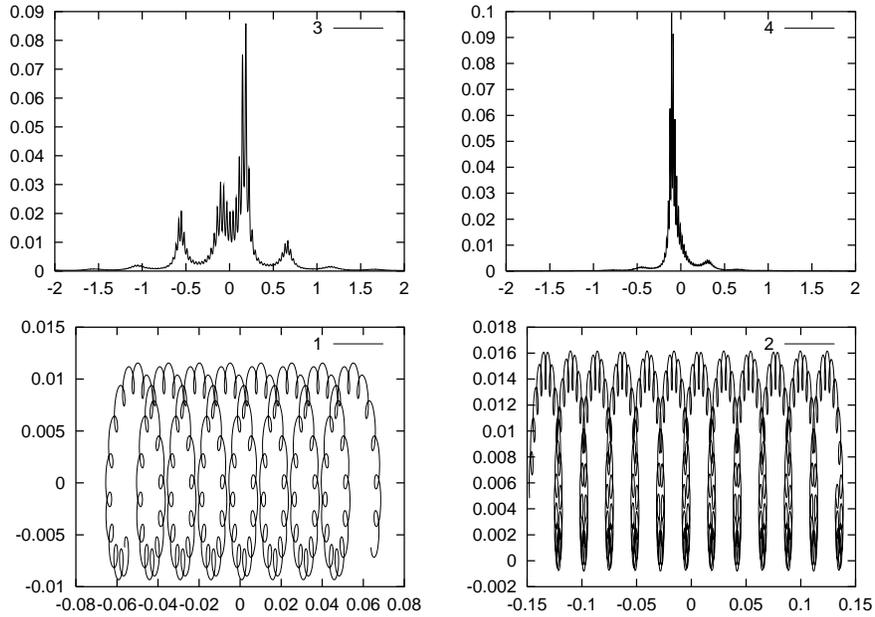, width= 12cm}}
  \caption{\em{Down : Label 1 and 2, two different
  bi-dimensional angle Spinning, Precessing Gamma Jet ring patterns toward the
   detector at the  origin ($0,0$).
   Up: Label 3-4, the consequent X, $\gamma$ intensity time evolution
   signals derived by ICS formula
   and characteristic beaming as in the text.}}
 \end{center}
\end{figure}

\section{Conclusions}

GRBs and SGRs  are persistent blazing flashes from light-house
thin $\gamma$ Jet spinning  in multi-precessing (binary,
precession, nutation)  mode. These Jets are originated by NSs or
BH in binary system or disk powered by infall matter: the Jet is
not a single explosive event even in GRB they are powered at
maximal output during SN event.The Jet power is comparable at its
peak  the $\gamma$ Jet has a chain of progenitor identities: it
is born in most SN and or BH birth and it is very probably
originated by very collimated primary muon pairs at GeVs-TeVs
energies. These muons could cross the dense target matter around
the SN explosions. These muon progenitors might be  themselves
secondary relics of pion decays or even by more transparent beamed
ultra-high energy neutrino Jet originated (by hadronic and pion
showering) near the NS or BH. The relativistic muons  decay in
flight in electron pairs is itself source of GeVs relativistic
pairs whose Inverse Compton Scattering with nearby thermal photon
is the final source of the observed  hard $X$ - $\gamma$ Jet. The
relativistic morphology of the Jet and its multi-precession is
the source of the puzzling complex $X$-$\gamma$ spectra signature
of GRBs and SGRs. Its inner internal Jet contain, following the
relativistic Inverse Compton Scattering,  hardest and rarest
beamed GeVs-MeVs photons (as the rarest EGRET GRB940217 one) but
its external Jet cones are dressed by softer and softer photons.
This   onion like multi Jets is not totally axis symmetric: it
doesn't appear on front as a concentric ring serial; while
turning and spraying around it is deformed (often) into an
elliptical off-axis concentric rings preceded by the internal
Harder center leading to a common Hard to Soft GRBs (and SGRs)
train signal. In our present model and simulation this internal
effect has been  here neglected without any major consequence. The
complex variability of GRBs and SGRs are simulated successfully
by the equations and the consequent geometrical beamed Jet
blazing leading to the observed $X-\gamma$ signatures. As shown
in fig 5 the slightly different precessing configurations could
easely mimic the wide morphology of GRBs as well as the
surprising rare X-ray precursor shown in Fig.1-4 above.

\section{Acknowledgment}
The author wish to thank Pier Giorgio De Sanctis Lucentini for
the valuable support and discussion.

\end{document}